# Transient Conjugate Free Convection from a Vertical Flat Plate in a Porous Medium Subjected to a Sudden Change in Surface Heat Flux

Jian-Jun SHU[*] and I. Pop

*School of Mechanical & Aerospace Engineering, Nanyang Technological University,
50 Nanyang Avenue, Singapore 639798*

---

[*] Corresponding author.
E-mail address: mjjshu@ntu.edu.sg (J.-J. Shu).

**Abstract**

The paper presents a theoretical study using the Kármán-Pohlhausen method for describing the transient heat exchange between the boundary-layer free convection and a vertical flat plate embedded in a porous medium. The unsteady behavior is developed after the generation of an impulsive heat flux step at the right-hand side of the plate. Two cases are considered according to whether the plate has a finite thickness or no thickness. The time and space evolution of the interface temperature is evidenced.

## 1. Introduction

The importance of heat transfer phenomena associated with free convection in porous media is well known. Interest in this phenomenon has been motivated by such diverse engineering problems as geothermal energy extraction, storage of nuclear waste material, ground water flows, pollutant dispersion in aquifers and packed-bed reactors, to mention just a few applications [1].

Owing to its fundamental and practical importance, the conjugate coupling heat transfer between a free convection flow and a vertical flat plate of finite thickness embedded in a porous medium has received particular attention [2-7]. Various approaches were used to deal with the difficulties associated with the simultaneous solution of the flow and thermal boundary layers and the longitudinal and transversal heat conduction in the solid plate. Despite the existing results in the open literature, they do not yet provide a complete description of this important problem, which has a bearing on many practical applications, particularly those related to energy conservation in buildings [8].

The point we wish to take up here is that of the transient conjugate free convection due to a vertical flat plate embedded in a porous medium. At a given time the right-hand side of the plate is suddenly subjected to a uniform heat flux, while the left-hand side of the plate is thermally insulated [9]. The present study is conducted in two phases: with finite thickness or without thickness of the plate, respectively. Analytical and numerical solutions are presented for all possible values of time and space evolution of the interface temperature.

## 2. Plate with Thickness

Consider unsteady free convection flow due to a semi-infinite vertical flat plate of finite thickness $a$ adjacent to a semi-infinite fluid-saturated porous medium. Initially, the whole system is at a temperature $T_\infty$, but subsequently the left-hand side of the plate is suddenly raised to, and held at a uniform heat flux $q_w$. The physical model and coordinate system is shown in Fig. 1. Assuming that the porous medium is isotropic and homogeneous and that the fluid is incompressible, the boundary layer and the Boussinesq approximations are invoked to obtain the following equations.

The equation of continuity
$$\frac{\partial u}{\partial x} + \frac{\partial v}{\partial y} = 0 \qquad (1)$$

The Darcy's Law
$$u = \frac{gK\beta}{\nu}(T_f - T_\infty) \qquad (2)$$

The equation of energy in the fluid-porous medium
$$\sigma \frac{\partial T_f}{\partial t} + u \frac{\partial T_f}{\partial x} + v \frac{\partial T_f}{\partial y} = \alpha_f \frac{\partial^2 T_f}{\partial y^2} \qquad (3)$$

and the equation of the heat transfer inside the solid plate
$$\frac{\partial T_s}{\partial t} = \alpha_s \frac{\partial^2 T_s}{\partial y^2} \qquad (4)$$

where ($x$, $y$) are the Cartesian coordinates along and normal to the plate, ($u$, $v$) are the velocity components in the ($x$, $y$) directions, $t$ is the time, $T_f$ and $T_s$ are the temperatures of the fluid-saturated porous medium and the solid plate, respectively, and $g$, $\beta$, $\nu$, $K$, $\alpha_f$, $\alpha_s$ and $\sigma$ are the physical constants. Equations (1)-(4) are subject to the following initial and boundary conditions.

For the fluid-porous medium ($y \geq 0$)
$$u = v = 0, \quad T_f = T_\infty \quad \text{at } t = 0 \text{ or } x = 0$$
$$v = 0 \quad \text{on } y = 0$$
$$u = 0, \quad T_f = T_\infty \quad \text{as } y \to \infty$$

for the solid ($-a \leq y \leq 0$)
$$T_s = T_\infty \quad \text{at } t = 0 \text{ or } x = 0$$
$$\frac{\partial T_s}{\partial y} = 0 \quad \text{on } y = -a$$

for the fluid-solid interface
$$T_f = T_s = T_p \quad \text{on } y = 0 \text{ and } t > 0$$
$$q_w = |q_s| = |q_f| = k_s \frac{\partial T_s}{\partial y} - k_f \frac{\partial T_f}{\partial y} \quad \text{on } y = 0 \text{ and } t > 0$$

where $T_p$ is the interface temperature and $k_f$ and $k_s$ are the thermal conductivities of the fluid and solid, respectively.

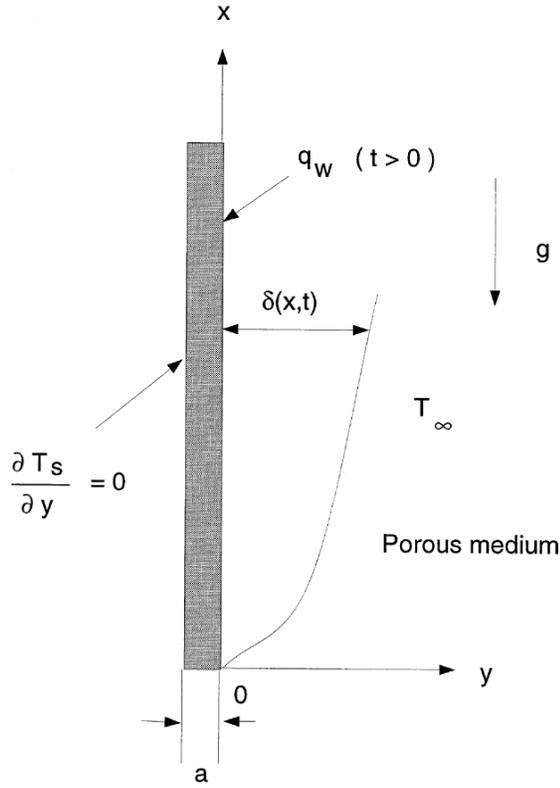

**Fig. 1.** Physical model and coordinate system

Further, a second-order Kármán-Pohlhausen temperature profile [10-15] is assumed in the fluid-porous medium to have

$$\theta_f = \theta_p \left(1 - \frac{y}{\delta}\right)^2, \quad \theta_s = \frac{1}{2k}\left(1 - \frac{2\theta_p}{\delta}\right)y^2 + \frac{1}{k}\left(1 - \frac{2\theta_p}{\delta}\right)y + \theta_p$$

where $\delta$ is the boundary-layer thickness and

$$\theta_f = \frac{k_f}{aq_w}(T_f - T_\infty), \quad \theta_s = \frac{k_f}{aq_w}(T_s - T_\infty), \quad k = \frac{k_s}{k_f}.$$

To obtain the integral form of the governing equations for transient conjugate free convection in a vertical porous layer, equations (1)-(4) are integrated across the boundary layer to yield

$$\frac{\Gamma}{3}\frac{\partial}{\partial t}(\delta\theta_p) + \frac{1}{5}\frac{\partial}{\partial x}(\delta\theta_p^2) = \frac{2\theta_p}{\delta}, \quad \frac{\partial}{\partial t}\left(\theta_p + \frac{2\theta_p}{3k\delta}\right) = \frac{1}{k}\left(1 - \frac{2\theta_p}{\delta}\right)$$

subject to

$$\delta = \theta_p = 0 \text{ at } t = 0 \text{ or } x = 0,$$

where $\Gamma = \frac{\sigma\alpha_s}{\alpha_f}$. They are hyperbolic sets of partial quasi-linear differential equations, which have two characteristic curves. By using the method of characteristics, the equations of direction of the characteristics are

$$dx = 0 \text{ and } \frac{\Gamma}{3}\left(\frac{4}{3} + k\delta\right)dx = \frac{1}{5}\theta_p(2 + k\delta)dt$$

so that the wave speed in the porous medium is

$$\frac{9(2 + k\delta)\theta_p}{5\Gamma(4 + 3k\delta)}.$$

The interface temperature distribution $\theta_p$ is illustrated in Fig. 2 to show that although the value of $\theta_p$ increases continuously with both in $t$ and $x$, its slope exhibits a discontinuity at whose value depends on $t$, $\Gamma$ and $k$. This discontinuity suggests a sudden change in the heat transfer characteristics that can be attributed to the presence of an essential singularity in the governing equations [16-23]. It is also noticed that $\theta_p$ increases continuously with time and approaches for large time the corresponding steady state value.

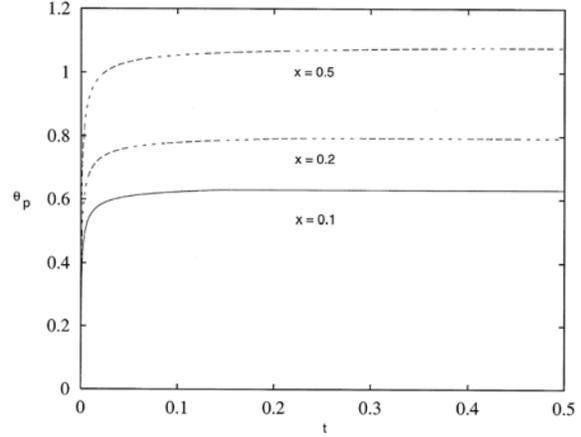

**Fig. 2.** Interface temperature for plate with thickness

### 3. Plate without Thickness

The same assumptions and restrictions are considered as in the previous section, but with a semi-infinite flat plate without thickness ($a = 0$). In this configuration, there are no conduction phenomena. It is noticed that equation has only one characteristic curve and the wave speed now is

$$\frac{9\theta_p}{10}.$$

The same behaviour of the temperature distribution $\theta_p$ can also be seen in Fig. 3.

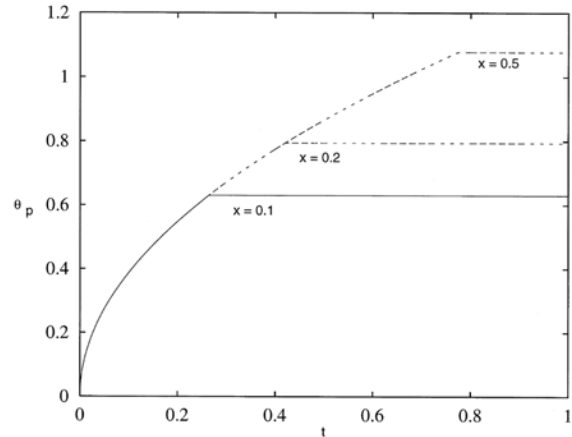

**Fig. 3.** Interface temperature for plate without thickness

**References**

[1] Shu J-J, Pop I. Inclined wall plumes in porous media. Fluid Dynamics Research 1997;21(4):303-317.

[2] Vynnycky M, Kimura S. Conjugate free convection due to a vertical plate in a porous medium. International Journal of Heat and Mass Transfer 1994;37(2):229-236.

[3] Vynnycky M, Kimura S. Transient conjugate free convection due to a vertical plate in a porous medium. International Journal of Heat and Mass Transfer 1995;38(2):219-231.

[4] Pop I, Lesnic D, Ingham DB. Conjugate mixed convection on a vertical surface in a porous medium. International Journal of Heat and Mass Transfer 1995;38(8):1517-1525.

[5] Lesnic D, Ingham DB, Pop I. Conjugate free convection from a horizontal surface in a porous medium. Zeitschrift für Angewandte Mathematik und Mechanik 1995;75(9):715-722.

[6] Pop I, Merkin JH. Conjugate free convection on a vertical surface in a saturated porous medium. Fluid Dynamics Research 1995;16(2-3):71-86.

[7] Higuera FJ, Pop I. Conjugate natural convection heat transfer between two porous media separated by a vertical wall. International Journal of Heat and Mass Transfer 1997;40(1):123-129.

[8] Treviño C, Méndez F, Higuera FJ. Heat transfer across a vertical wall separating two fluids at different temperatures. International Journal of Heat and Mass Transfer 1996;39(11):2231-2241.

[9] Shu J-J, Pop I. Thermal interaction between free convection and forced convection along a vertical conducting wall. Heat and Mass Transfer 1999;35(1):33-38.

[10] Cheng P. Convective heat transfer in porous layers by integral methods. Letters in Heat and Mass Transfer 1978;5(5):243-252.

[11] Shu J-J, Wilks G. Heat transfer in the flow of a cold, two-dimensional vertical liquid jet against a hot, horizontal plate. International Journal of Heat and Mass Transfer 1996;39(16):3367-3379.

[12] Shu J-J. Microscale heat transfer in a free jet against a plane surface. Superlattices and Microstructures 2004;35(3-6):645-656.

[13] Shu J-J, Wilks G. Heat transfer in the flow of a cold, axisymmetric vertical liquid jet against a hot, horizontal plate. Journal of Heat Transfer-Transactions of the ASME 2008;130(1):012202.

[14] Shu J-J, Wilks G. Heat transfer in the flow of a cold, two-dimensional draining sheet over a hot, horizontal cylinder. European Journal of Mechanics B-Fluids 2009;28(1):185-190.

[15] Shu J-J, Wilks G. Heat transfer in the flow of a cold, axisymmetric jet over a hot sphere. Journal of Heat Transfer-Transactions of the ASME 2013;135(3):032201.

[16] Yang K-T. Remarks on transient laminar free convection along a vertical plate. International Journal of Heat and Mass Transfer 1966;9(5):511-513.

[17] Nanbu K. Limit of pure conduction for unsteady free convection on a vertical plate. International Journal of Heat and Mass Transfer 1971;14(9):1531-1534.

[18] Pop I, Cheng P. The growth of a thermal layer in a porous medium adjacent to a suddenly heated semi-infinite horizontal surface. International Journal of Heat and Mass Transfer 1983;26(10):1574-1576.

[19] Cheng P, Pop I. Transient free convection about a vertical flat plate embedded in a porous medium. International Journal of Engineering Science 1984;22(3):253-264.

[20] Shu J-J, Wilks G. An accurate numerical method for systems of differentio-integral equations associated with multiphase flow. Computers & Fluids 1995;24(6):625-652.

[21] Shu J-J, Wilks G. Mixed-convection laminar film condensation on a semi-infinite vertical plate. Journal of Fluid Mechanics 1995;300:207-229.

[22] Shu J-J, Pop I. On thermal boundary layers on a flat plate subjected to a variable heat flux. International Journal of Heat and Fluid Flow 1998;19(1):79-84.

[23] Shu J-J. Laminar film condensation heat transfer on a vertical, non-isothermal, semi-infinite plate. Arabian Journal for Science and Engineering 2012;37(6):1711-1721.